\documentclass[english,pre,a4paper,aps,reprint]{revtex4-2}
\usepackage[T1]{fontenc}
\usepackage[latin9]{inputenc}
\usepackage{amsmath}
\usepackage{amssymb}
\usepackage{graphicx}

\makeatletter
\usepackage{natbib}

\makeatother

\usepackage{babel}
\begin{document}
\title{Apparent pathologies in stochastic entropy production in the thermalisation
of an open two-level quantum system }
\author{Jonathan Dexter and Ian J. Ford }
\affiliation{Department of Physics and Astronomy, University College London, Gower
Street, London WC1E 6BT, U.K.}
\begin{abstract}
We investigate the entropic consequences of the relaxation of an open
two-level quantum system towards a thermalised statistical state,
using a framework of quantum state diffusion with a minimal set of
raising and lowering Lindblad operators. We demonstrate that thermalisation
starting from a general state is accompanied by a persistent non-zero
mean rate of change of the environmental component of stochastic entropy
production. This thermodynamic signature can be associated with the
purification of the reduced density matrix $\rho$ of the randomly
evolving state, to be contrasted with the impurity of the more often
considered ensemble average of $\rho$. The system adopts stationary
statistics, with zero stochastic entropy production, once purity has
been achieved. However, we show that apparent pathological mathematical
difficulties in the computation of stochastic entropy production emerge
if $\rho$ is represented using a certain set of coordinates, though
these can be removed by choosing a different set. We conclude that
frameworks for modelling open quantum systems must be carefully selected
to provide satisfactory thermodynamic as well as dynamic behaviour.
\end{abstract}
\maketitle

\section{Introduction}

The second law of thermodynamics emerges when the world is described
at a coarse grained level, but governed by underlying equations of
motion with a sufficient degree of deterministic chaos \citep{Ford-book2013}.
Entropy increase corresponds to a growth in subjective uncertainty
regarding the adopted configuration of the world in such a situation,
and this applies both to classical and to quantum systems \citep{matos2022,Clarke23}.
To illustrate the latter, we study the evolution of an open two-level
quantum system using a minimal framework of stochastic dynamics that
represents the effects of coupling to an underspecified environment.
We treat the reduced density matrix of the system as a set of physical
coordinates that undergo Brownian motion, and concern ourselves with
the mechanical and thermodynamic approach to a thermalised state.
The stochastic entropy production that accompanies the evolution is
a measure of its thermodynamic irreversibility and identifies an arrow
of time \citep{seifert2008stochastic}.

We employ a framework of quantum state diffusion \citep{percival1998quantum}
where the system dynamics are modelled using Itô processes: Markovian
stochastic differential equations (SDEs) with continuous solutions.
The associated stochastic entropy production can also be described
using an SDE \citep{spinney2012use,spinney2012entropy}. We find several
apparently pathological features in the stochastic entropy production
and resolve them. In Section \ref{sec:An-open-quantum} we encounter
persistent stochastic entropy production accompanying thermalisation
of the system by the environment, suggesting that a stationary statistical
state is reached only in infinite time. We link this behaviour with
asymptotic purification under the chosen dynamics. Furthermore, relaxation
of a pure system towards the thermalised state can appear to be burdened
with mathematical difficulties in the form of singularities in stochastic
entropy production. We investigate and resolve such peculiarities
in Section \ref{sec:Pathologies-in-stochastic}. Our conclusions are
given in Section \ref{sec:Conclusions}. 

Appendix \ref{sec:Stochastic-entropy-production-1} provides a discussion
of the basis for computing stochastic entropy production, and Appendix
\ref{sec:Reduction-of-a-1} is a summary of how to do so in situations
characterised by restricted diffusion \citep{Dexter23}. Appendix
\ref{sec:System-contribution-to} explores some unexpected mathematical
features that arise in the averaging of SDEs, and how they can lead
to additional terms in the mean system component of stochastic entropy
production.

\section{Thermalising a two-level open quantum system \label{sec:An-open-quantum}}

\subsection{System dynamics}

A stochastic Lindblad equation for the evolution of the reduced density
matrix $\rho$ of an open system can be constructed using Lindblad
operators that describe the pseudorandom influence of the environment
on the system \citep{jacobs2014quantum,matos2022,Clarke23}. Such
an approach allows us to model the evolution of $\rho$ given an underspecified
state of the environment: a concept familiar in classical statistical
mechanics but less so in quantum mechanics. No projective measurement
is imposed. Environmental interactions drive quantum state diffusion
such that $\rho$ follows a continuous Brownian trajectory.

We consider a two-level quantum system with environmental couplings
characterised by the raising and lowering Lindblad operators given
by
\begin{equation}
c_{+}=\begin{pmatrix}0 & 0\\
1 & 0
\end{pmatrix},\qquad c_{-}=\begin{pmatrix}0 & 1\\
0 & 0
\end{pmatrix},\label{eq: twolevelrandl}
\end{equation}
expressed in a basis of eigenstates of the $\sigma_{z}$ Pauli matrix.
The stochastic Lindblad equation is

\begin{equation}
\begin{split}d\rho & =\left(c_{+}\rho c_{+}^{\dagger}-\frac{1}{2}\rho c_{+}^{\dagger}c_{+}-\frac{1}{2}c_{+}^{\dagger}c_{+}\rho\right)dt\\
 & +\left(\rho c_{+}^{\dagger}+c_{+}\rho-{\rm Tr}\left[\left(c_{+}+c_{+}^{\dagger}\right)\rho\right]\rho\right)dW_{1}\\
 & +\left(c_{-}\rho c_{-}^{\dagger}-\frac{1}{2}\rho c_{-}^{\dagger}c_{-}-\frac{1}{2}c_{-}^{\dagger}c_{-}\rho\right)dt\\
 & +\left(\rho c_{-}^{\dagger}+c_{-}\rho-{\rm Tr}\left[\left(c_{-}+c_{-}^{\dagger}\right)\rho\right]\rho\right)dW_{2},
\end{split}
\label{eq: drho2lind}
\end{equation}
where $dW_{1,2}$ are independent Wiener increments \citep{matos2022,Clarke23,Dexter23}.
Expressing $\rho$ in terms of components of the coherence or Bloch
vector $\boldsymbol{r}=(r_{x},r_{y},r_{z})\equiv(x,y,z)$, we write

\begin{equation}
\rho=\frac{1}{2}\begin{pmatrix}1+z & x-iy\\
x+iy & 1-z
\end{pmatrix},\label{eq:rho}
\end{equation}
namely $\rho=\frac{1}{2}\left(\mathbb{I}+\boldsymbol{r}\cdot\boldsymbol{\sigma}\right)$
in terms of Pauli matrices $\sigma_{i}$. After some manipulation
we obtain three SDEs describing the dynamics of the system: 

\begin{equation}
\begin{pmatrix}dx\\
dy\\
dz
\end{pmatrix}=\begin{pmatrix}-x\\
-y\\
-2z
\end{pmatrix}dt+\begin{pmatrix}1-x^{2}-z & 1-x^{2}+z\\
-xy & -xy\\
x\left(1-z\right) & -x\left(1+z\right)
\end{pmatrix}\begin{pmatrix}dW_{1}\\
dW_{2}
\end{pmatrix},\label{eq: sdevector}
\end{equation}
which take the form 
\begin{equation}
dr_{i}=A_{i}(\boldsymbol{r},t)dt+\sum_{j}B_{ij}(\boldsymbol{r},t)dW_{j},\label{eq: sde form}
\end{equation}
in terms of a vector $\boldsymbol{A}$ and matrix $\boldsymbol{B}$. 

\begin{figure}
\centering{}\includegraphics[width=0.9\linewidth]{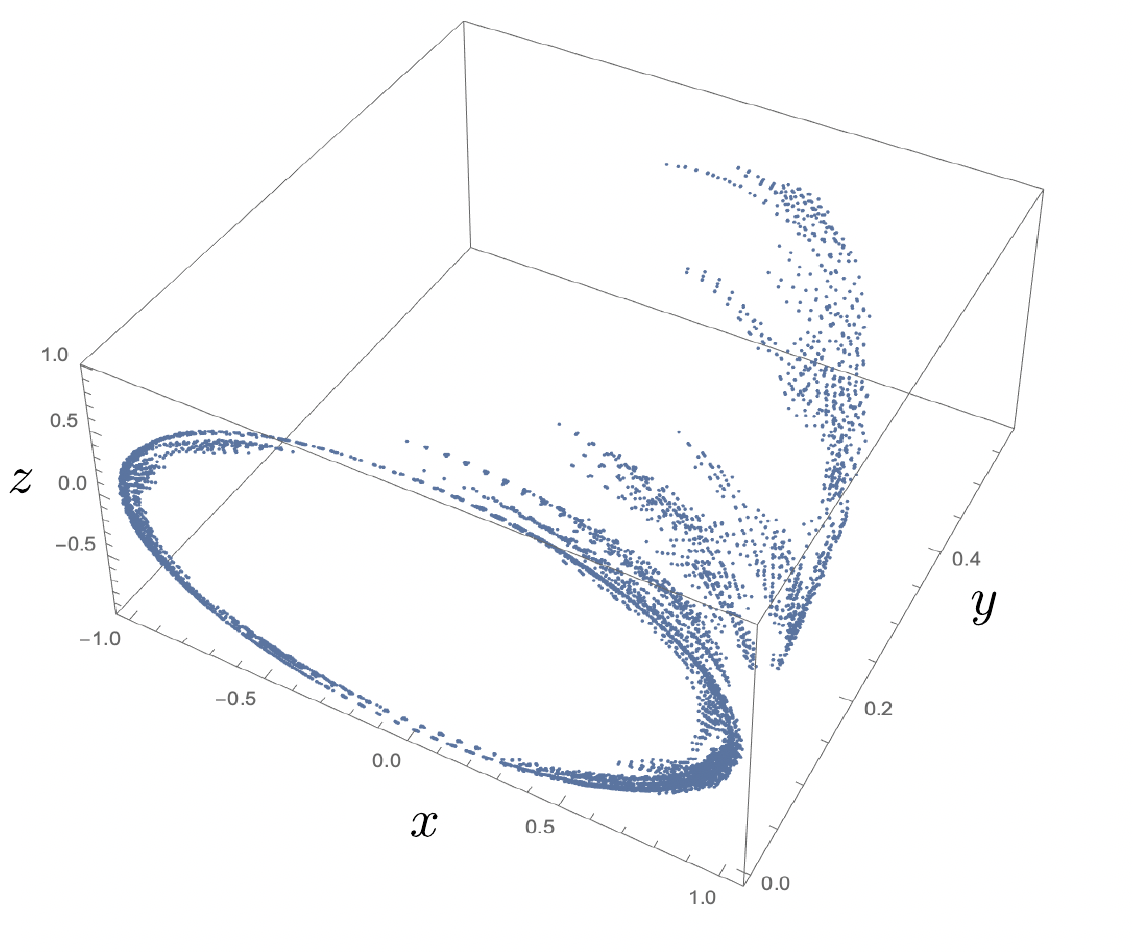} \caption{Example evolution of the coherence vector $\boldsymbol{r}=(x,y,z)$
governed by Eq. (\ref{eq: sdevector}). The trajectory was based on
10000 steps with timestep $dt=10^{-3}$ and initiated at the point
$x=y=z=0.5$. The coherence vector appears to be confined to an ellipsoid
and to be driven towards a circle of unit radius in the $x$-$z$
plane, corresponding to evolution towards purity.}
\label{fig: ellipsoid}
\end{figure}

\begin{figure}[h]
\centering \includegraphics[width=1\columnwidth]{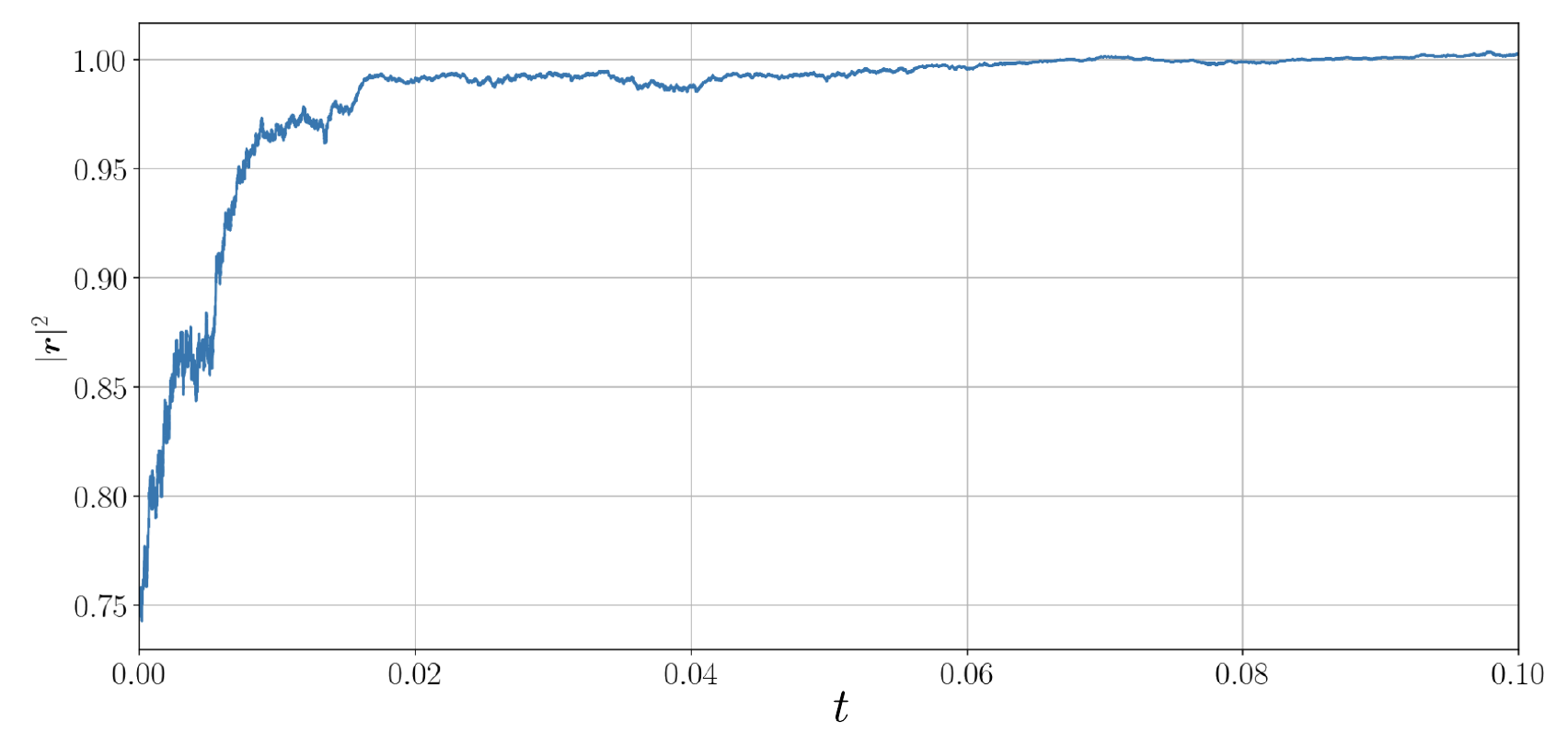} \caption{Evolution of the magnitude of the coherence vector $\boldsymbol{r}$
under the dynamics described by Eq. (\ref{eq: sdevector}). We see
that $|\boldsymbol{r}|^{2}$ is driven towards unity, which is synonymous
with evolution towards purity, $P=1$. The trajectory was initiated
at the point $x=y=z=0.5$ and constructed using a timestep of $dt=10^{-5}$.}
\label{fig:spherical purity}
\end{figure}

The number of coordinates (three) is larger than the number of noise
terms (two), suggesting that a reduced description exists involving
only two coordinates. Indeed an example of the evolution of $\boldsymbol{r}$
in its phase space in Fig. \ref{fig: ellipsoid} suggests the motion
is constrained to lie on the surface of an ellipsoid lying inside
the Bloch sphere defined by $\vert\boldsymbol{r}\vert=r=1$. The purity
of the system $P={\rm Tr}\rho^{2}=\frac{1}{2}\left(1+r^{2}\right)$
evolves according to
\begin{equation}
dP=4\left(1-P\right)\left(1-x^{2}\right)dt+4x\left(1-P\right)\left(dW_{1}+dW_{2}\right),\label{eq:purity}
\end{equation}
suggesting that purification $P\to1$ takes place as $t\to\infty$.
This is demonstrated in Fig. \ref{fig:spherical purity}.

\subsection{Stochastic entropy production}

The incremental stochastic entropy production associated with the
stochastic dynamics, which measures the irreversibility of the evolution,
is usually separated into system and environmental components as follows:
\begin{equation}
d\Delta s_{{\rm tot}}=d\Delta s_{\text{sys}}+d\Delta s_{\text{env}},\label{eq:sde for Delta s}
\end{equation}
with $d\Delta s_{\text{sys}}=-d\ln p(\boldsymbol{r},t)$ where $p(\boldsymbol{r},t)$
is the probability density function (pdf) characterising the stochastic
dynamical variables, governed by an appropriate Fokker-Planck equation.
The average of the system component $d\Delta s_{\text{sys}}$ over
all possible trajectories is related to the incremental change in
Gibbs entropy of the system: $d\langle\Delta s_{{\rm sys}}\rangle=dS_{G}$,
to which boundary terms should be added in certain circumstances \citep{matos2022},
as shown in Appendix \ref{sec:System-contribution-to}. We shall encounter
a case where boundary contributions are important in Section \ref{sec:Pathologies-in-stochastic}. 

An SDE for the evolution of the environmental component $d\Delta s_{\text{env}}$
of stochastic entropy production \citep{spinney2012use} is given
in Appendix \ref{sec:Stochastic-entropy-production-1}. A significant
problem in computing $d\Delta s_{\text{env}}$ using this SDE is that
the determinant of the diffusion matrix in the Fokker-Planck equation
can be zero, making it impossible to define an inverse matrix needed
in Eq. (\ref{eq: senvbig}). This pathology, however, can be overcome,
as we now demonstrate using procedures summarised in Appendix \ref{sec:Reduction-of-a-1}. 

\subsection{Environmental stochastic entropy production in $x,z$ coordinates\label{subsec:Environmental-stochastic-entropy-1} }

The confinement of the trajectory to an ellipsoid in Fig. \ref{fig: ellipsoid}
is a consequence of the existence of a constant function of the motion 

\begin{equation}
f\left(x,y,z\right)=\frac{1}{y^{2}}\left(1-x^{2}-z^{2}\right),\label{eq: ellipsoid}
\end{equation}
which evolves by Itô's lemma according to
\begin{equation}
\begin{split}df & =\sum_{i}\frac{\partial f}{\partial r_{i}}dr_{i}+\sum_{i,j}\frac{\partial^{2}f}{\partial r_{i}\partial r_{j}}D_{ij}dt,\end{split}
\label{eq: itoreduced-1}
\end{equation}
where elements of the diffusion matrix are $D_{ij}=\frac{1}{2}\sum_{k}B_{ik}B_{jk}$
and the matrix $\boldsymbol{B}$ is defined in Eq. (\ref{eq: sde form}).
Using Eq. (\ref{eq: sdevector}) we obtain

\begin{equation}
\boldsymbol{D}=\begin{pmatrix}x^{4}-2x^{2}+z^{2}+1 & xy\left(x^{2}-1\right) & xz\left(x^{2}-2\right)\\
xy\left(x^{2}-1\right) & x^{2}y^{2} & x^{2}yz\\
xz\left(x^{2}-2\right) & x^{2}yz & x^{2}\left(z^{2}+1\right)
\end{pmatrix},\label{eq: d matrix}
\end{equation}
and by insertion into Eq. (\ref{eq: itoreduced-1}) we find that $df=0$,
 demonstrating $f$ to be a constant of the motion. 

\begin{figure}
\centering \includegraphics[width=1\columnwidth]{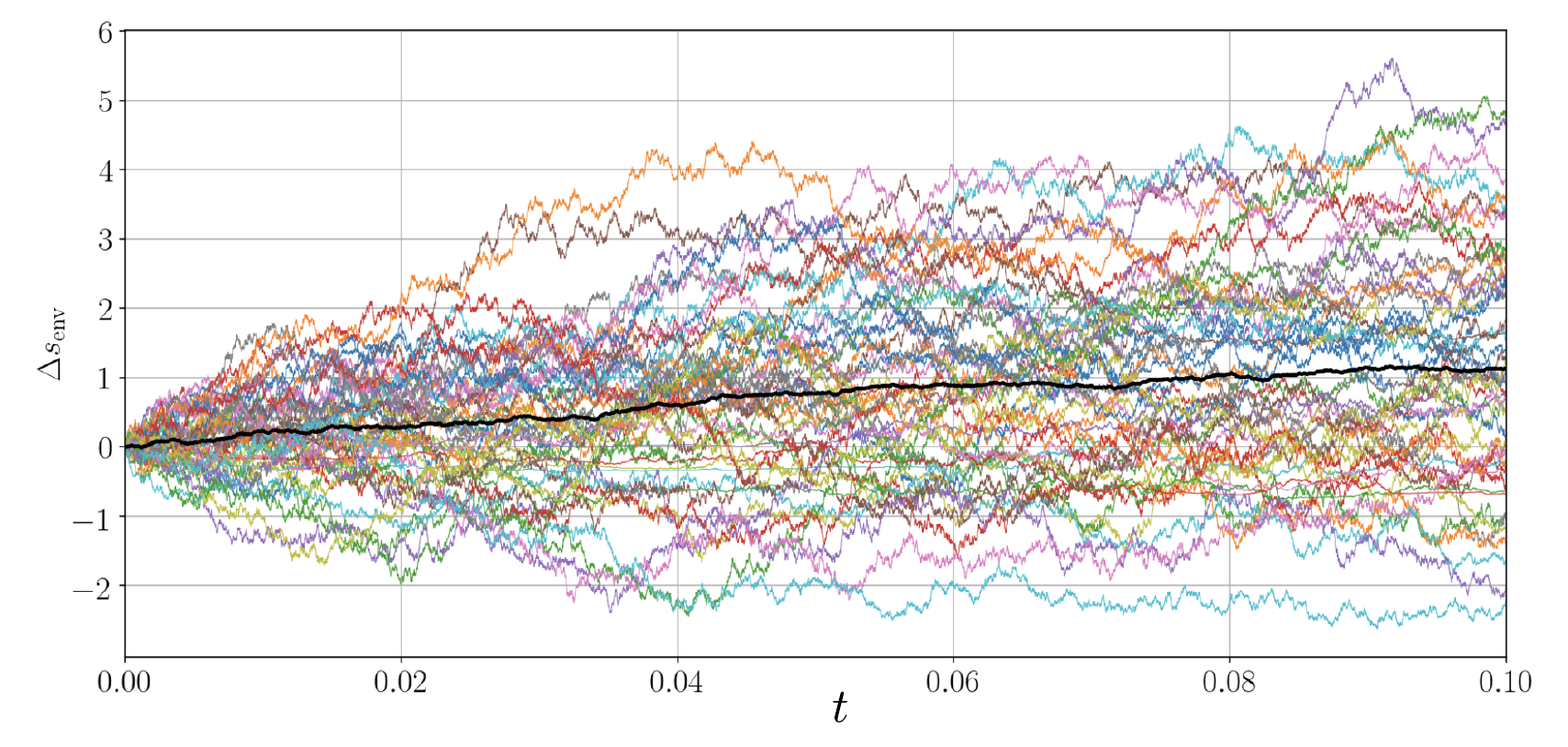}
\caption{Environmental component of stochastic entropy production computed
using Eq. (\ref{eq: senvbig}) for 50 realisations of the dynamics
described by Eq. (\ref{eq:51}). The black line represents the ensemble
mean. Each trajectory was initiated at the point $x=z=0.5$ and was
evaluated using a timestep of $dt=10^{-5}$.}
\label{fig:xz entropy}
\end{figure}

The constancy of $f$ creates a situation characterised by restricted
diffusion \citep{Dexter23}. The dynamics in Eq. (\ref{eq: sdevector})
express evolution in an $N=3$ dimensional phase space driven by $M=2$
noise terms. Following the arguments summarised in Appendix \ref{sec:Reduction-of-a-1},
we identify the single $\left(N-M=1\right)$ eigenvector of $\boldsymbol{D}$
with an eigenvalue equal to zero, finding this to be $\boldsymbol{\alpha}=\begin{pmatrix}\frac{x}{z}, & \frac{1-x^{2}-z^{2}}{yz}, & 1\end{pmatrix}^{\mathsf{T}}$.
The existence of a `null' eigenvector allows us to remove one of the
SDEs from the dynamics. If we eliminate the evolution of coordinate
$y$ the dynamics are then expressed according to

\begin{equation}
\begin{pmatrix}dx\\
dz
\end{pmatrix}=\begin{pmatrix}-x\\
-2z
\end{pmatrix}dt+\begin{pmatrix}1-x^{2}-z & 1-x^{2}+z\\
x\left(1-z\right) & -x\left(1+z\right)
\end{pmatrix}\begin{pmatrix}dW_{1}\\
dW_{2}
\end{pmatrix},\label{eq:51}
\end{equation}
and the resulting reduced diffusion matrix is given by

\begin{equation}
\boldsymbol{D}_{{\rm red}}=\begin{pmatrix}x^{4}-2x^{2}+z^{2}+1 & xz\left(x^{2}-2\right)\\
xz\left(x^{2}-2\right) & x^{2}\left(z^{2}+1\right)
\end{pmatrix},\label{eq:52}
\end{equation}
which is non-singular. We can now compute trajectories for $d\Delta s_{\text{env}}$
using Eq. (\ref{eq: senvbig}) with Eq. (\ref{eq:51}) and the reduced
diffusion matrix (\ref{eq:52}). Figure \ref{fig:xz entropy} illustrates
50 realisations of the dynamics together with the average evolution.

The increase in $d\Delta s_{\text{env}}$ with time, on average, seems
to be a pathology: we might expect entropy production to cease as
the system becomes thermalised. However, the dynamics are incapable
of achieving thermalisation in finite time. The purity approaches
unity only as $t\to\infty$, according to Eq. (\ref{eq:purity}),
and as this evolution continues, stochastic entropy production will
persist. The situation is analogous to behaviour found in a two-level
system undergoing measurement \citep{Clarke23}. Thermalisation is
only achieved asymptotically, and the production of stochastic entropy
continues to measure progress in this direction. This resolves the
issue, but in the next Section we turn to a more puzzling matter.

\section{Pathologies in stochastic entropy production for a pure state\label{sec:Pathologies-in-stochastic} }

Consider the situation where purity has been established such that
$y\to0$ and $x^{2}\to1-z^{2}$. The coherence vector then evolves
randomly on a circle of unit radius in the $x$-$z$ plane and the
stochastic dynamics and stochastic entropy production for the further
relaxation of the pure state involve the evolution of a single coordinate
$z$. We give this situation our attention in order to resolve an
apparent pathology in the stochastic entropy production. 

Inserting the limits for $x$ and $y$ into Eq. (\ref{eq: sdevector})
leads to an SDE 
\begin{equation}
dz=-2zdt+\left(2\left(1-z^{4}\right)\right)^{1/2}dW,\label{eq:b1-1}
\end{equation}
describing the thermalisation of a pure state of the two-level system
through the Lindblads $c_{\pm}$. Note that in the absence of a system
Hamiltonian the thermal state will be characterised by equal occupational
probabilities of the two levels, each represented by coherence vectors
at the north and south poles of the Bloch sphere, or $z=\pm1$. The
mean value of $z$ will tend asymptotically to zero as a consequence.
Note that if the Lindblad operators were unequally weighted, for example
using $\left(1-\gamma/2\right)^{1/2}c_{+}$ and $\left(1+\gamma/2\right)^{1/2}c_{-}$
where $\gamma$ is a constant, the SDE would read
\begin{equation}
dz=-\left(\gamma+2z\right)dt+\left(2\left(1-z^{2}\right)\left(1+\gamma z+z^{2}\right)\right)^{1/2}dW,\label{eq:finite T}
\end{equation}
which has a stationary state with $\langle z\rangle=-\gamma/2$.

However, there are difficulties in computing the stochastic entropy
production associated with the dynamics of Eq. (\ref{eq:b1-1}), which
we now identify.

\subsection{Environmental component}

Referring to Eq. (\ref{eq:b1-1}) and Appendix \ref{sec:Stochastic-entropy-production-1},
the environmental component of stochastic entropy production takes
the form
\begin{eqnarray}
d\Delta s_{{\rm env}} & = & \frac{A^{{\rm irr}}}{D}dz+\frac{\partial A^{{\rm irr}}}{\partial z}dt-\frac{1}{D}\frac{\partial D}{\partial z}dz\nonumber \\
 &  & -\frac{A^{{\rm irr}}}{D}\frac{\partial D}{\partial z}dt-\frac{\partial^{2}D}{\partial z^{2}}dt+\frac{1}{D}\left(\frac{\partial D}{\partial z}\right)^{2}dt,\qquad\label{dstot expression-1}
\end{eqnarray}
with $A^{{\rm irr}}=-2z$ and $D=1-z^{4}$. Inserting $dD/dz=-4z^{3}$
and $d^{2}D/dz^{2}=-12z^{2}$, we obtain
\begin{align}
d\Delta s_{{\rm env}} & =2z\frac{\left(2z^{2}-1\right)}{\left(1-z^{4}\right)^{1/2}}2^{1/2}dW\nonumber \\
 & +\frac{2}{1-z^{4}}\left(-1-7z^{4}+8z^{2}+2z^{6}\right)dt,\label{eq:b2-1}
\end{align}
 which has singularities at $z=\pm1$, potentially creating difficulties
in evaluating $d\Delta s_{{\rm env}}$ numerically.

\subsection{System component\label{subsec:System-component}}

\begin{figure}
\includegraphics[width=1\columnwidth]{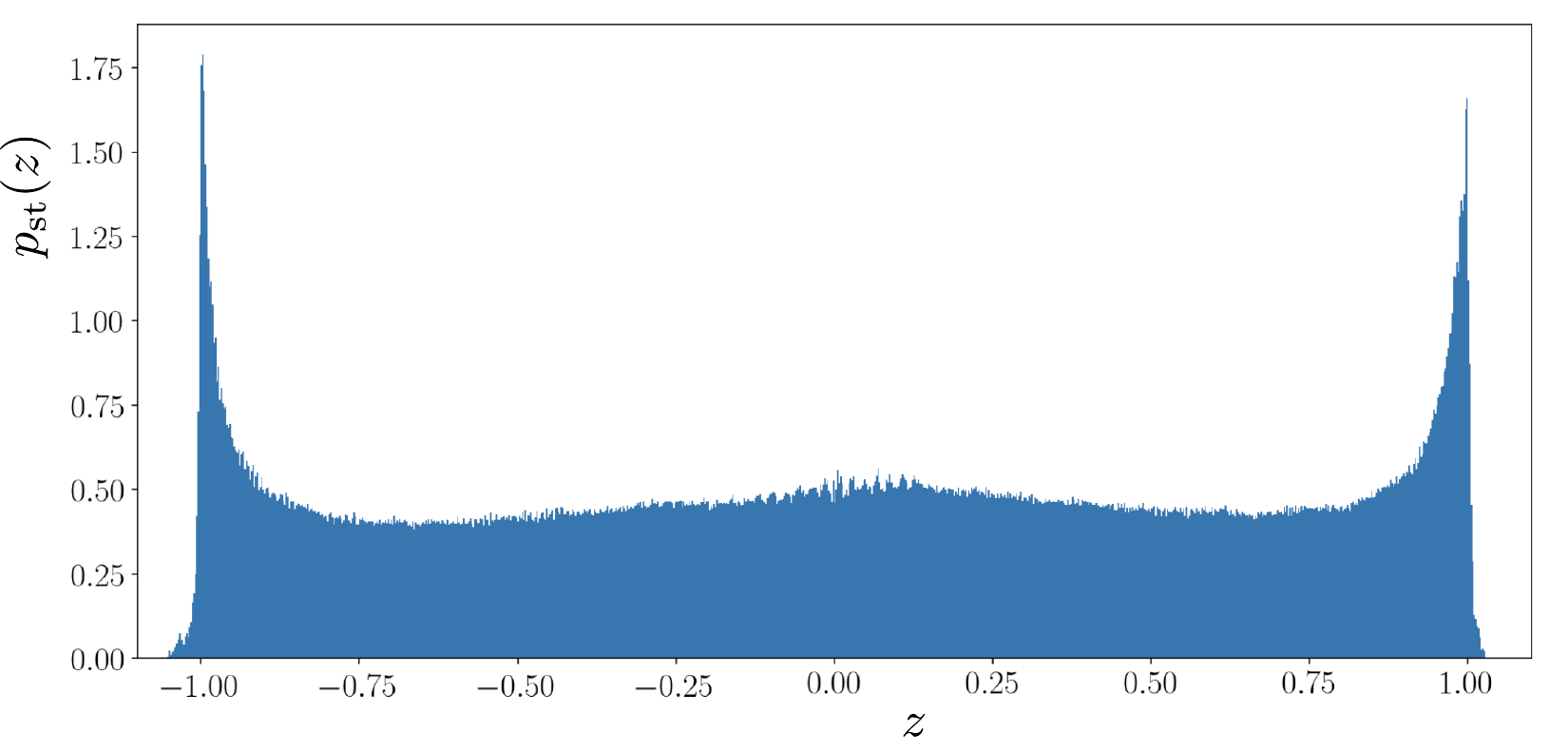} \caption{Histogram illustrating the stationary pdf of $z$ generated from a
solution to Eq. (\ref{eq: sdevector}) using $10^{7}$ timesteps of
size $dt=10^{-3}$. The trajectory was initiated at the point $x=y=z=0.5$
and the bin size is $10^{-4}$.}
\label{fig:hist}
\end{figure}

Further problems emerge when we consider the system component of stochastic
entropy production. This is apparent from the histogram of values
of $z$ in Fig. \ref{fig:hist}, obtained from a single trajectory
generated using the set of SDEs in Eq. (\ref{eq: sdevector}). The
dynamics clearly allow $z$ to take values $\pm1$, suggesting that
the singularities in the environmental component of stochastic entropy
production identified in Eq. (\ref{eq:b2-1}) might be encountered.

We can derive the stationary pdf analytically to illustrate this further.
The Fokker-Planck equation associated with Eq. (\ref{eq:b1-1}) is
\begin{equation}
\frac{\partial p(z,t)}{\partial t}=-\frac{\partial J}{\partial z},\label{eq:fpe pure}
\end{equation}
where the current is
\begin{equation}
J=-2zp-\frac{\partial}{\partial z}\left(\left(1-z^{4}\right)p\right).\label{eq:z1}
\end{equation}
The stationary pdf in $-1\le z\le1$ is given by

\begin{equation}
p_{{\rm st}}(z)\propto\frac{1}{\left(1-z^{4}\right)^{1/2}}\frac{1}{1+z^{2}},\label{eq:z3}
\end{equation}
which possesses singularities at $z=\pm1$, consistent with the numerical
results in Fig. \ref{fig:hist}. 

Note that the stationary expectation value $\int_{-1}^{1}zp_{{\rm st}}(z)dz=0$
is compatible with the asymptotic solution to the evolution $d\langle z\rangle=-2\langle z\rangle dt$
obtained by averaging the SDE (\ref{eq:b1-1}) over the noise and
employing $\langle dW\rangle=0$, where angled brackets denote stochastic
averages. In Appendix \ref{sec:System-contribution-to} we show, however,
that this result need not hold if the product of diffusion coefficient
and pdf at the boundaries does not vanish: an extra term contributes
such that 
\begin{equation}
d\langle z\rangle=-2\langle z\rangle dt-\left[Dp\right]_{z=-1}^{z=1}dt.\label{eq: z4}
\end{equation}
In this case we have $Dp_{{\rm st}}\propto(1-z^{4})^{1/2}/(1+z^{2})$,
which vanishes at $z=\pm1$. Assuming, not unreasonably, that this
feature is preserved when the system is nonstationary, the usual outcome
of the stochastic averaging of Eq. (\ref{eq:b1-1}) emerges.

Boundary terms also appear in the evolution of the mean system component
of stochastic entropy production. According to Eq. (\ref{eq:a14}),
the relationship between this quantity and the change in Gibbs entropy
$S_{G}$ is:
\begin{align}
d\langle\Delta s_{{\rm sys}}\rangle & =\frac{dS_{G}(t)}{dt}dt-\left[J\ln p\right]_{z=-1}^{z=1}dt-\left[D\frac{\partial p}{\partial z}\right]_{z=-1}^{z=1}dt\nonumber \\
- & \left[D\left(\frac{\partial\ln p}{\partial z}\right)^{2}P(\Delta s_{{\rm sys}},t)\right]_{\Delta s_{{\rm sys}}^{{\rm min}}}^{\Delta s_{{\rm sys}}^{{\rm max}}}dt.\label{eq:a14-2}
\end{align}
The first additional term vanishes since $J=0$ at the boundaries
and the third can be simplified using $P(\Delta s_{{\rm sys}},t)d\Delta s_{{\rm sys}}=pdz$.
In the stationary state both the second and third terms depend on
the expression
\begin{align}
D\frac{dp_{{\rm st}}}{dz} & \propto\left(1-z^{4}\right)\Big(2z^{3}\left(1-z^{4}\right)^{-3/2}\left(1+z^{2}\right)^{-1}\nonumber \\
 & -2z\left(1-z^{4}\right)^{-1/2}\left(1+z^{2}\right)^{-2}\Big),\label{eq:pathology}
\end{align}
which is singular at $z=\pm1$. We therefore expect difficulties to
arise in computing the system stochastic entropy production, in addition
to those already encountered in the evaluation of $\Delta s_{{\rm env}}$.
But are these genuine problems?

\subsection{Removal of pathological behaviour}

The pathologies may be removed by a simple change of coordinate frame.
Defining $\theta=\cos^{-1}z$ and using Itô's lemma, we find that
the dynamics of the pure state may also be described by the SDE

\begin{equation}
d\theta=\frac{1}{2}\sin2\theta dt+\left(2\left(1+\cos^{2}\theta\right)\right)^{1/2}dW,\label{eq:c1}
\end{equation}
instead of Eq. (\ref{eq:b1-1}) for $z$. The diffusion coefficient
is $D=1+\cos^{2}\theta$. The stationary pdf in $\theta$ is
\begin{equation}
p_{{\rm st}}(\theta)\propto\left(1+\cos^{2}\theta\right)^{-3/2},\label{eq:c2}
\end{equation}
and so $Dp_{{\rm st}}\propto\left(1+\cos^{2}\theta\right)^{-1/2}$
such that the boundary term $\left[Dp_{{\rm st}}\right]_{\theta=0}^{\theta=\pi}$
vanishes. As a consequence, the averaging of Eq. (\ref{eq:c1}) proceeds
along usual lines, giving $d\langle\theta\rangle=\frac{1}{2}\langle\sin2\theta\rangle dt$.

Furthermore, we have
\begin{equation}
D\frac{dp_{{\rm st}}}{d\theta}\propto\frac{\sin2\theta}{1+\cos^{2}\theta},\label{eq:c3}
\end{equation}
which is zero at $\theta=0$ and $\pi$. As a consequence, the boundary
terms in the evolution of mean system stochastic entropy production
in the stationary state vanish. It is not unreasonable to expect that
even in a nonstationary situation the evolution should satisfy
\begin{align}
d\langle\Delta s_{{\rm sys}}\rangle & =\frac{dS_{G}(t)}{dt}dt,\label{eq:c4}
\end{align}
again in line with what is typically assumed.

Finally, the evolution of environmental stochastic entropy production
can be identified using the SDE for $\theta$:
\begin{equation}
d\Delta s_{{\rm env}}=\frac{6+18\cos2\theta+3\sin^{2}2\theta}{4\left(1+\cos^{2}\theta\right)}dt+\frac{3\sin2\theta}{\sqrt{2}\left(1+\cos^{2}\theta\right)}dW,\label{eq:c5}
\end{equation}
in which there are once again no singularities at the boundaries.
We conclude that all the pathologies disappear when an appropriate
coordinate system is used to describe the dynamics and thermodynamics. 

\section{Conclusions\label{sec:Conclusions}}

The main purpose in presenting this study is to identify interpretational
and mathematical issues concerning stochastic entropy production in
a minimal model of the thermalisation of a two-level quantum system,
and then to resolve them. The treatment of open system dynamics as
a continuous Brownian motion of a physical density matrix, a quantum
state diffusion, is perhaps unfamiliar and the consideration of stochastic
entropy production as the outcome of an Itô process particularly so.
The mean value of stochastic entropy production is a measure of the
change in subjective uncertainty in the adopted state of the world
given the employment of a coarse grained model of its evolution. We
need to address any apparent pathologies in the calculation of stochastic
entropy production that might cast doubt on its use in this role.

Our conclusions are as follows. Thermalisation of a two-level system
can be achieved through coupling to a coarse grained environment by
way of raising and lowering operators. Such an interaction is not
the same as quantum measurement, but nevertheless the minimal model
we use purifies the reduced density matrix asymptotically in time,
and this brings about a persistent production of stochastic entropy.
This is not a pathology and indeed it seems very reasonable that thermal
interactions should disentangle a system from its environment, creating
greater uncertainty in the state of the world as the more classical,
less correlated system behaviour emerges. Note that an impure or mixed
reduced density matrix, such as the Gibbs state often used to describe
a thermalised system, is to be interpreted in the framework of quantum
state diffusion as the \emph{average} state of an ensemble of pure
reduced density matrices. The mixed reduced density matrix does not
necessarily imply entanglement.

Furthermore, if the two-level system is thermalised starting from
a disentangled or pure state, the stochastic entropy production can
appear to be pathological in both the environmental and system components.
It turns out, however, that such a conclusion is an artefact of a
particular choice of coordinate system and a simple transformation
of variables can remove the mathematical difficulties. Stochastic
entropy production should not depend on choice of coordinates, but
some are more suitable than others. We note as a corollary that the
separate system and environmental components of stochastic entropy
production \emph{are }coordinate frame dependent. Gibbs entropy, for
example, expresses uncertainty of adopted system state in a particular
coordinate phase space. 

In conclusion, we have presented arguments to support the use of stochastic
entropy production in open quantum system dynamics in a manner similar
to its employment in classical situations. Its role and the implications
are just the same, and we suggest, just as profound.

\appendix

\section{Stochastic entropy production\label{sec:Stochastic-entropy-production-1}}

We consider a set of coordinates $\boldsymbol{x}\equiv(x_{1},x_{2},\cdots)$
that specify the configuration of a system, and model their evolution
using Itô processes:

\begin{equation}
dx_{i}=A_{i}(\boldsymbol{x})dt+\sum_{j}B_{ij}(\boldsymbol{x})dW_{j},\label{eq:a100-1}
\end{equation}
where the $dW_{j}$ are independent Wiener increments. The Fokker-Planck
equation for the pdf $p(\boldsymbol{x},t)$ is
\begin{equation}
\frac{\partial p}{\partial t}=-\sum_{i}\frac{\partial}{\partial x_{i}}\left(A_{i}p\right)+\frac{\partial}{\partial x_{i}\partial x_{j}}\left(D_{ij}p\right),\label{eq: fpe}
\end{equation}
in terms of a diffusion matrix $\boldsymbol{D}(\boldsymbol{x})=\frac{1}{2}\boldsymbol{B}(\boldsymbol{x})\boldsymbol{B}(\boldsymbol{x})^{\mathsf{T}}$.
The stochastic entropy production of the system, and its environment,
associated with the stochastic dynamics is given by 
\begin{equation}
d\Delta s_{{\rm tot}}=d\Delta s_{\text{sys}}+d\Delta s_{\text{env}},\label{eq: dstot}
\end{equation}
with $d\Delta s_{\text{sys}}=-d\ln p(\boldsymbol{x},t)$. \noindent\begin{widetext}
If we define $A_{i}^{{\rm irr}}(\boldsymbol{x})=\frac{1}{2}\left[A_{i}(\boldsymbol{x})+\varepsilon_{i}A_{i}(\boldsymbol{\varepsilon}\boldsymbol{x})\right]$
and $A_{i}^{{\rm rev}}(\boldsymbol{x})=\frac{1}{2}\left[A_{i}(\boldsymbol{x})-\varepsilon_{i}A_{i}(\boldsymbol{\varepsilon}\boldsymbol{x})\right]$
\citep{SpinneyFord12b} where $\varepsilon_{i}=1$ for variables $x_{i}$
with even parity under time reversal symmetry (for example position)
and $\varepsilon_{i}=-1$ for variables with odd parity (for example
velocity), with $\boldsymbol{\varepsilon}\boldsymbol{x}$ representing
$(\varepsilon_{1}x_{1},\varepsilon_{2}x_{2},\cdots)$, it may be shown
that \citep{spinney2012use} 
\begin{equation}
\begin{split} & d\Delta s_{\text{env}}=-\sum_{i}\frac{\partial A_{i}^{\text{rev}}(\boldsymbol{x})}{\partial x_{i}}dt+\sum_{i,j}\Biggl\{\frac{D_{ij}^{-1}(\boldsymbol{x})}{2}\left(A_{i}^{\text{irr}}(\boldsymbol{x})dx_{j}+A_{j}^{\text{irr}}(\boldsymbol{x})dx_{i}\right)\\
 & -\frac{D_{ij}^{-1}(\boldsymbol{x})}{2}\left(\left(\sum_{n}\frac{\partial D_{jn}(\boldsymbol{x})}{\partial x_{n}}\right)dx_{i}+\left(\sum_{m}\frac{\partial D_{im}(\boldsymbol{x})}{\partial x_{m}}\right)dx_{j}\right)-\frac{D_{ij}^{-1}(\boldsymbol{x})}{2}\left(A_{i}^{\text{rev}}(\boldsymbol{x})A_{j}^{\text{irr}}(\boldsymbol{x})+A_{j}^{\text{rev}}(\boldsymbol{x})A_{i}^{\text{irr}}(\boldsymbol{x})\right)dt\\
 & +\frac{D_{ij}^{-1}(\boldsymbol{x})}{2}\left(A_{j}^{\text{rev}}(\boldsymbol{x})\left(\sum_{m}\frac{\partial D_{im}(\boldsymbol{x})}{\partial x_{m}}\right)+A_{i}^{\text{rev}}(\boldsymbol{x})\left(\sum_{n}\frac{\partial D_{jn}(\boldsymbol{x})}{\partial x_{n}}\right)\right)dt\\
 & +\frac{1}{2}\sum_{k}\Biggl[D_{ik}(\boldsymbol{x})\frac{\partial}{\partial x_{k}}\left(D_{ij}^{-1}(\boldsymbol{x})A_{j}^{\text{irr}}(\boldsymbol{x})\right)+D_{jk}(\boldsymbol{x})\frac{\partial}{\partial x_{k}}\left(D_{ij}^{-1}(\boldsymbol{x})A_{i}^{\text{irr}}(\boldsymbol{x})\right)\\
 & -D_{ik}(\boldsymbol{x})\frac{\partial}{\partial x_{k}}\left(D_{ij}^{-1}(\boldsymbol{x})\left(\sum_{n}\frac{\partial D_{jn}(\boldsymbol{x})}{\partial x_{n}}\right)\right)-D_{jk}(\boldsymbol{x})\frac{\partial}{\partial x_{k}}\left(D_{ij}^{-1}(\boldsymbol{x})\left(\sum_{m}\frac{\partial D_{im}(\boldsymbol{x})}{\partial x_{m}}\right)\right)\Biggr]dt\Biggr\}.
\end{split}
\label{eq: senvbig}
\end{equation}
\end{widetext}

\section{Dimensional reduction of a singular diffusion matrix\label{sec:Reduction-of-a-1}}

Consider a system evolving stochastically in an $N$ dimensional phase
space under the influence of $M$ noise terms such that the $\boldsymbol{B}$
matrix in Eq. (\ref{eq:a100-1}) is $N\times M$ and the diffusion
matrix $\boldsymbol{D}$ in the Fokker-Planck equation (\ref{eq: fpe})
is $N\times N$. The phase space is described by a set of $N$ coordinates
$\left\{ x_{1},x_{2},...,x_{N}\right\} $, and we conjecture that
there exist $L=N-M$ functions of these coordinates that are invariant
under the dynamics, or more generally, evolve without noise. These
allow us to remove $L$ coordinates from the $N$ dimensional phase
space and consider evolution in a reduced $M$ dimensional phase space.
We call the $M$ coordinates $\left(\left\{ x_{m}\right\} \right)$
of the reduced phase space \emph{dynamical variables} and the remaining
$L$ coordinates $\left(\left\{ x_{l}\right\} \right)$ \emph{spectator
variables}. The $L$ constraints allow us to write the spectator variables
as functions of the dynamical variables: $\left\{ x_{l}\left(\left\{ x_{m}\right\} \right)\right\} $.
The diffusion matrix in the reduced phase space will be $M\times M$
and non-singular allowing us to use Eq. (\ref{eq: senvbig}) to compute
the stochastic entropy production associated with the evolution \citep{Dexter23}. 

If $f(\boldsymbol{x})$ is a constant of the motion, we can regard
$\boldsymbol{\nabla}f$ as a `null' eigenvector of $\boldsymbol{D}$
with zero eigenvalue. Furthermore, since the infinitesimal displacement
$d\boldsymbol{x}$ lies on the hypersurface of $f$ to which the evolution
is confined, we also have $\boldsymbol{\nabla}f\cdot d\boldsymbol{x}=0$.
Combining these we write

\begin{equation}
\sum_{m=1}^{M}\alpha_{km}dx_{m}+\sum_{l=1}^{L}\alpha_{kl}dx_{l}=0,\label{eq: split}
\end{equation}
where $\alpha_{km}$ and $\alpha_{kl}$ are the dynamical and spectator
components of the $k$th null eigenvector of $\boldsymbol{D}$. The
$\alpha_{km}$ are elements of a rectangular $L\times M$ matrix $Q_{km}$
and the $\alpha_{kl}$ are elements of a square $L\times L$ matrix
$P_{kl}$ so that $P_{kl}dx_{l}=-Q_{km}dx_{m}$ and

\begin{equation}
\begin{split}dx_{l} & =-P_{lk}^{-1}Q_{km}dx_{m}=R_{lm}dx_{m},\end{split}
\label{eq: rmatrix}
\end{equation}
where summations over repeated indices are implied and $R_{lm}$ is
a component of the $L\times M$ matrix $\boldsymbol{R}$. Next we
write 

\begin{equation}
dD_{ij}=\sum_{m=1}^{M}\frac{\partial D_{ij}}{\partial x_{m}}dx_{m}+\sum_{l=1}^{L}\frac{\partial D_{ij}}{\partial x_{l}}dx_{l},\label{eq: chainrule}
\end{equation}
and substituting from Eq. (\ref{eq: rmatrix}) we obtain the derivative
of $D_{ij}$ with respect to the dynamical coordinate $x_{m}$ in
the form

\begin{equation}
\frac{dD_{ij}}{dx_{m}}=\frac{\partial D_{ij}}{\partial x_{m}}+\sum_{l}\frac{\partial D_{ij}}{\partial x_{l}}R_{lm}.\label{eq: derivatives}
\end{equation}
Equation (\ref{eq: senvbig}) can now be used to compute stochastic
entropy production based on the dynamics of the $M$ dynamical variables
with derivatives modified according to Eq. (\ref{eq: derivatives}).
Additional discussion of the approach is given elsewhere \citep{Dexter23}.

\section{System component of mean stochastic entropy production\label{sec:System-contribution-to}}

It is typically assumed that the average of the $d\Delta s_{{\rm sys}}=-d\ln p(\boldsymbol{x},t)$
term in the expression for the incremental stochastic entropy production
corresponds to the change in system Gibbs entropy, but this is not
necessarily the case, as we now show. 

Consider the SDE for a function $g(\boldsymbol{x})$ of the stochastic
variables $\boldsymbol{x}$:
\begin{equation}
dg=\hat{A}(\boldsymbol{x})dt+\hat{B}(\boldsymbol{x})dW,\label{eq:a1}
\end{equation}
with pdf $P(g,t)$ satisfying
\begin{equation}
\frac{\partial P(g,t)}{\partial t}=-\frac{\partial\hat{J}}{\partial g},\label{eq:a2}
\end{equation}
in terms of a probability current $\hat{J}$ given by
\begin{equation}
\hat{J}=\hat{A}P-\frac{1}{2}\frac{\partial}{\partial g}\left(\hat{B}^{2}P\right).\label{eq:a3}
\end{equation}
Defining a stochastic average of $g$ as $\langle g\rangle_{t}=\int gP(g,t)dg$,
we consider the difference $d\langle g\rangle$ between $\langle g\rangle_{t+dt}$
and $\langle g\rangle_{t}$:
\begin{align}
d\langle g\rangle & =\int g\left(P(g,t+dt)-P(g,t)\right)dg=\int g\frac{\partial P}{\partial t}dgdt\nonumber \\
 & =-\int g\frac{\partial\hat{J}}{\partial g}dgdt=-\left[g\hat{J}\right]dt+\int\hat{J}\,dgdt,\label{eq:a5}
\end{align}
where the square brackets indicate boundary terms. Since $\hat{J}=0$
at the boundaries in the absence of probability current sources and
sinks, the first term vanishes and so
\begin{align}
d\langle g\rangle & =\int\left(\hat{A}P-\frac{1}{2}\frac{\partial}{\partial g}\left(\hat{B}^{2}P\right)\right)dgdt\nonumber \\
 & =\langle\hat{A}\rangle_{t}dt-\frac{1}{2}\left[\hat{B}^{2}P\right]dt,\label{eq:a6}
\end{align}
where another boundary term has appeared. Note that the average in
the first term on the right hand side can be written in two forms:
$\langle\hat{A}\rangle_{t}=\int\hat{A}(\boldsymbol{x})P(g,t)dg=\int\hat{A}(\boldsymbol{x})p(\boldsymbol{x},t)d\boldsymbol{x}$.
The second version involves the pdf $p(\boldsymbol{x},t)$ over the
stochastic variables $\boldsymbol{x}$, as specified by the Fokker-Planck
equation
\begin{equation}
\frac{\partial p}{\partial t}=-\sum_{j}\frac{\partial J_{j}}{\partial x_{j}},\label{eq:a6a}
\end{equation}
associated with the dynamics $dx_{i}=A_{i}dt+\sum_{j}B_{ij}dW_{j}.$
The probability current is
\begin{equation}
J_{j}=A_{j}p-\sum_{i}\frac{\partial}{\partial x_{i}}\left(D_{ij}p\right),\label{eq:a6b}
\end{equation}
where $D_{ij}=\frac{1}{2}\sum_{k}B_{ik}B_{jk}$. 

The first term on the right hand side of Eq. (\ref{eq:a6}) would
perhaps be expected to be the only contribution after averaging Eq.
(\ref{eq:a1}), but it is augmented by a term that depends on the
values of $\hat{B}^{2}P$ at the boundaries of the domain of $g$.
This extra term could vanish but it is not guaranteed to do so. 

Consider, for example, a single stochastic variable evolving as $dx=adt+bdW$:
the stochastic average satisfies $d\langle x\rangle/dt=\langle a\rangle-[Dp]$
where $D=b^{2}/2$. In many cases, the pdf vanishes at the boundaries,
or for periodic boundary conditions the contributions might cancel.
It is only in certain cases that further attention is required. An
explicit case involves $a(x)=x$ and $b(x)=x^{2}$ in the domain $0\le x\le\infty$.
The stationary pdf is $p_{{\rm st}}(x)=4\pi^{-1/2}x^{-4}\exp(-1/x^{2})$
and the diffusion coefficient is $D=x^{4}/2.$ Hence $Dp_{{\rm st}}$
vanishes at $x=0$ but is equal to $2/\pi^{1/2}$ at $x=\infty$.
We should hence model the evolution of $\langle x\rangle$ according
to $d\langle x\rangle/dt=\langle x\rangle-2/\pi^{1/2}$, which is
consistent with a non-zero mean of $x$ as $t\to\infty$, as suggested
by the stationary pdf. The boundary term is crucial here.

Let us now analyse the evolution of the system stochastic entropy
production, $g(\boldsymbol{x})=\Delta s_{{\rm sys}}=-\ln p(\boldsymbol{x},t)$.
We write
\begin{align}
d\Delta s_{{\rm sys}} & =-\frac{\partial\ln p}{\partial t}dt-\sum_{j}\frac{\partial\ln p}{\partial x_{j}}dx_{j}-\sum_{ij}D_{ij}\frac{\partial^{2}\ln p}{\partial x_{i}\partial x_{j}}dt\nonumber \\
 & =\hat{A}dt+\sum_{ij}\hat{B}_{ij}dW_{j}=\hat{A}dt+\hat{B}dW,\label{eq:a7}
\end{align}
with
\begin{equation}
\hat{A}=-\frac{\partial\ln p}{\partial t}-\sum_{j}\frac{\partial\ln p}{\partial x_{j}}A_{j}-\sum_{ij}D_{ij}\frac{\partial^{2}\ln p}{\partial x_{i}\partial x_{j}},\label{eq:a8}
\end{equation}
and
\begin{equation}
\hat{B}_{ij}=-\frac{\partial\ln p}{\partial x_{i}}B_{ij},\label{eq:a9}
\end{equation}
such that 
\begin{align}
\hat{B} & =\left(\sum_{ijk}\hat{B}_{kj}\hat{B}_{ij}\right)^{1/2}=\left(\sum_{ijk}\frac{\partial\ln p}{\partial x_{k}}\frac{\partial\ln p}{\partial x_{i}}B_{kj}B_{ij}\right)^{1/2}\nonumber \\
 & =\left(2\sum_{ik}\frac{\partial\ln p}{\partial x_{k}}\frac{\partial\ln p}{\partial x_{i}}D_{ki}\right)^{1/2}.\label{eq:a10}
\end{align}
The mean system stochastic entropy production therefore satisfies
\begin{equation}
d\langle\Delta s_{{\rm sys}}\rangle=\langle\hat{A}\rangle_{t}dt-\frac{1}{2}\left[\hat{B}^{2}P(\Delta s_{{\rm sys}},t)\right]dt,\label{eq:a11}
\end{equation}
where $P$ is the pdf of $\Delta s_{{\rm sys}}$. $\hat{B}$ is a
function of the stochastic variables $\boldsymbol{x}$ and the second
term is evaluated at the boundary values of $\Delta s_{{\rm sys}}$
using the appropriate values of $\boldsymbol{x}$.

We further develop the first term on the right hand side of Eq. (\ref{eq:a11}),
writing
\begin{equation}
\langle\hat{A}\rangle_{t}=\int\hat{A}(\boldsymbol{x}(\Delta s_{{\rm sys}}))P(\Delta s_{{\rm sys}},t)d\Delta s_{{\rm sys}},\label{eq:a12}
\end{equation}
and noting that for a given $\Delta s_{{\rm sys}}$ there are potentially
several corresponding values of $\boldsymbol{x}$, each of which contributes
to the integrand. It is better to cast the integral in terms of the
$\boldsymbol{x}$ instead of $\Delta s_{{\rm sys}}$, namely $\langle\hat{A}\rangle_{t}=\int\hat{A}(\boldsymbol{x})p(\boldsymbol{x},t)d\boldsymbol{x}$
and employing integration limits for $\boldsymbol{x}$ instead of
$\Delta s_{{\rm sys}}$. Continuing the development we find:
\begin{align}
\langle\hat{A}\rangle_{t} & =\int\left(-\frac{\partial\ln p}{\partial t}-\sum_{j}\frac{\partial\ln p}{\partial x_{j}}A_{j}-\sum_{ij}D_{ij}\frac{\partial^{2}\ln p}{\partial x_{i}\partial x_{j}}\right)pd\boldsymbol{x}\nonumber \\
= & \int\left(-\frac{\partial\ln p}{\partial t}p-\sum_{j}\left(A_{j}p-\frac{\partial\left(D_{ij}p\right)}{\partial x_{i}}\right)\frac{\partial\ln p}{\partial x_{j}}\right)d\boldsymbol{x}\nonumber \\
 & -\sum_{ij}\left[D_{ij}\frac{\partial\ln p}{\partial x_{j}}p\right]\nonumber \\
= & \int\left(-\frac{\partial\ln p}{\partial t}p+\sum_{j}\frac{\partial J_{j}}{\partial x_{j}}\ln p\right)d\boldsymbol{x}\nonumber \\
 & -\sum_{j}\left[J_{j}\ln p\right]-\sum_{ij}\left[D_{ij}\frac{\partial p}{\partial x_{j}}\right],\label{eq: A13}
\end{align}
reducing to
\begin{align}
\langle\hat{A}\rangle_{t} & =\int\left(-\frac{\partial\ln p}{\partial t}p-\frac{\partial p}{\partial t}\ln p\right)d\boldsymbol{x}\nonumber \\
 & -\sum_{j}\left[J_{j}\ln p\right]-\sum_{ij}\left[D_{ij}\frac{\partial p}{\partial x_{j}}\right]\nonumber \\
= & -\frac{d}{dt}\int p\ln p\,d\boldsymbol{x}-\sum_{j}\left[J_{j}\ln p\right]-\sum_{ij}\left[D_{ij}\frac{\partial p}{\partial x_{j}}\right].\label{eq:A11b-1}
\end{align}
Combining Eqs. (\ref{eq:a11}) and (\ref{eq:A11b-1}), and inserting
the Gibbs entropy $S_{G}(t)=-\int p(\boldsymbol{x},t)\ln p(\boldsymbol{x},t)\,d\boldsymbol{x}$,
we arrive at
\begin{align}
d\langle\Delta s_{{\rm sys}}\rangle & =\frac{dS_{G}(t)}{dt}dt-\sum_{j}\left[J_{j}\ln p\right]dt\nonumber \\
 & -\sum_{ij}\left[D_{ij}\frac{\partial p}{\partial x_{j}}\right]dt-\frac{1}{2}\left[\hat{B}^{2}P(\Delta s_{{\rm sys}},t)\right]dt.\label{eq:a14}
\end{align}
The increment in mean system stochastic entropy production corresponds
in part to the increment in Gibbs entropy $dS_{G}=\frac{dS_{G}}{dt}dt$,
but three boundary terms also contribute, the first two involving
the limits of the phase space of the $\boldsymbol{x}$ coordinates,
and the third the limits of the range of $\Delta s_{{\rm sys}}$.

\bibliography{ref}

\end{document}